\documentclass[%
 reprint,
 amsmath,amssymb,
prb,
]{revtex4-1}

\usepackage{amsmath,amssymb} 
\usepackage{physics} 
\usepackage{graphicx}
\usepackage{dcolumn}
\usepackage{bm}

\usepackage{soul}
\usepackage{colortbl}
\usepackage{hhline}

\usepackage{color}
\definecolor{blu}{rgb}{0,0,1}

\definecolor{grn}{rgb}{0,0.7,0}

\definecolor{ora}{rgb}{0.93,0.53,0.18}

\definecolor{pur}{rgb}{0.41,0.05,0.68}

\newcommand\numberthis{\addtocounter{equation}{1}\tag{\theequation}}

\begin{document}


\title{Dressing due to correlations strongly reduces the effect of electron-phonon coupling}

\author{Yau-Chuen Yam}
\author{George A. Sawatzky}
\author{Mona Berciu}
\affiliation{\!Department \!of \!Physics and Astronomy, \!University
  of\!  British Columbia, \!Vancouver, British \!Columbia,\! Canada,\!
 }

\affiliation{\!Stewart Blusson Quantum Matter \!Institute, \!University
  of British Columbia, \!Vancouver, British \!Columbia, \!Canada}

\date{\today}

\begin{abstract}
 We investigate the difference between the coupling of a bare carrier to phonons versus the coupling of a correlations-dressed quasiparticle to phonons, and show that latter may be weak even if the former is strong. Specifically, we analyze the effect of the hole-phonon coupling on the dispersion of the quasiparticle that forms when a single hole is doped into a cuprate layer. To model this, we start from the three-band Emery model supplemented by the Peierls modulation of the $p$-$d$ and $p$-$p$ hoppings due to the motion of O ions. We then project onto the strongly correlated $U_{dd}\rightarrow \infty$ limit where charge fluctuations are frozen on the Cu site. The resulting effective Hamiltonian describes the motion of a doped hole on the O sublattice, and its interactions with Cu spins and O phonons.  We show that even though the hole-phonon coupling is moderate to strong, it leads to only a very minor increase of  the quasiparticle's effective mass as compared to its mass  in the absence of coupling to phonons, consistent with a weak coupling to phonons of the correlations-dressed quasiparticle. We explain the reasons for this suppression, revealing why it is expected to happen in any systems with strong correlations.  
\end{abstract}

\pacs{Valid PACS appear here}
\maketitle

\section{\label{sec:intro}Introduction}

The study of many-body systems with both strong correlations and strong electron-phonon coupling is a formidable challenge, given that even understanding  the effects of only correlations and of only strong electron-phonon couplings is far from simple or complete. To simplify things, it is often customary to study model Hamiltonians that describe low-energy quasiparticles that are already correlations dressed, and couple these to phonons. In this work, we use the example of a cuprate layer doped with a single hole, to clarify the difference between the coupling to phonons of the bare carrier versus the correlations-dressed quasiparticle, and explain why the latter is generically much weaker than the former. 

It is well-known that cuprates  become high temperature superconductors upon hole doping, although the reason for this behavior is still under debate. Their common constituent feature, the  CuO$_2$ layer, is believed to be hosting the relevant electronic orbitals responsible for this phenomenology.\cite{Dam2003,Lee2006} However, despite decades of effort, it is still not clear what is the simplest model Hamiltonian that properly includes all the ingredients needed to understand the behavior of a doped CuO$_2$ layer.

\begin{figure}[t]
	\centering
	\includegraphics[width=0.75\columnwidth]{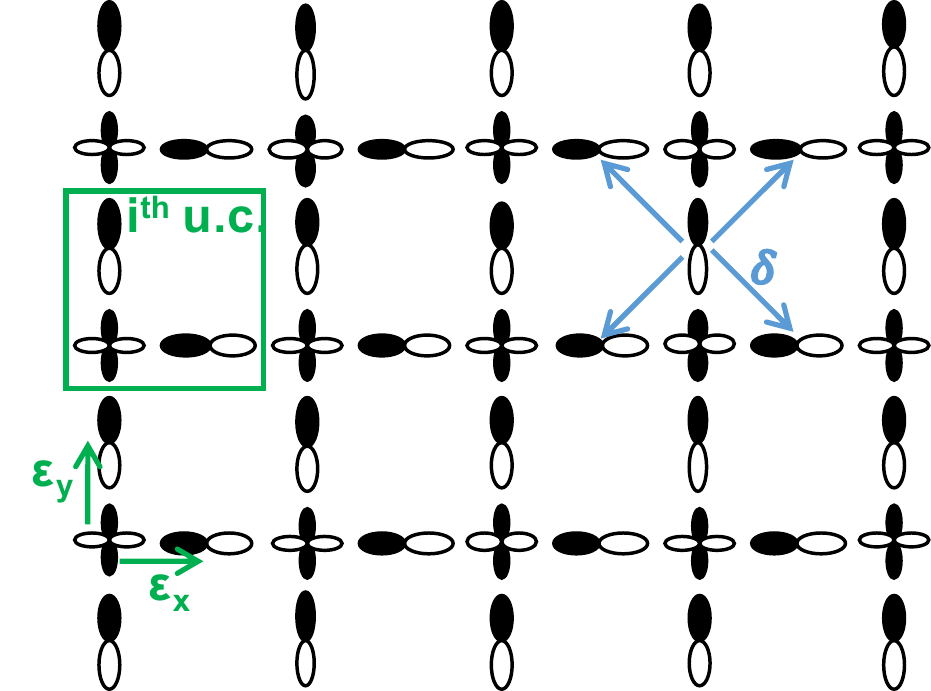}
	\caption{\label{lat_for_H} Sketch of the CuO$_2$ plane, with a unit cell (u.c.) comprising a Cu $3d_{x^2-y^2}$ valence orbital and O  $2p_x$ and  $2p_y$ ligand valence orbitals. We use $\epsilon_x,\epsilon_y$ (green arrows) to denote the displacement from the Cu to the two O in the same unit cell. The four vectors $\delta$ (blue arrows) are for the displacement from an O to its four adjacent O sites.
        }
\end{figure}

One of the well established starting points for describing the behavior of holes doped in a cuprate layer is  the three-band Emery model, involving  $3d_{x^2-y^2}$ Cu orbitals and the ligand $2p_{\sigma}$ O orbitals \cite{Eme1987,Esk1988}, as sketched in Fig. \ref{lat_for_H}. This model ignores other potentially important ingredients such as the  Cu $3d_z$ orbitals or the apical O, yet it is already too complicated to solve. Numerical simulations suffer from limitations of finite system sizes \cite{Bon2007}, issues from sign problems\cite{Loh1990,Mon2022}, lack of convergence at low-enough temperatures\cite{Sca1991}, etc., while a full analytical solution seems impossible. A well established way of simplifying the Emery model is to freeze charge fluctuations at the Cu sites to  one hole/Cu site (as is the case in the parent insulator) and describe them in terms of the spins of these Cu holes. The additional doped holes are moving in the O sublattice, as is known to be the case for this charge-transfer gap insulator.\cite{Zaa1985} Based on the resulting strong exchange interactions between the spin of a doped hole, located on an O,  and  the spins on its two sandwiching Cu sites, Emery and Reiter suggested a variational doped ground state known as the 3-spin polaron \cite{Eme1988}. This is a linear combination of singlets with the two Cu spins on either side of the hole,  with a relative phase that results in a ferromagnetic coupling between the two Cu spins.

The difficulties in dealing with either the full Emery model, or its simpler version with spins at the Cu sites and doped holes on the O,  has motivated efforts to simplify them even more. After Anderson \cite{And1987} pointed out that a single-band effective Hubbard Hamiltonian might suffice to describe the low-energy properties of a cuprate layer, Zhang and Rice \cite{Zha1988} mapped the simplified Emery model onto a one-band $t$-$J$ model (the strongly correlated limit of the Hubbard model) by projecting it onto the so-called Zhang-Rice singlets (ZRS). The ZRS is a different variational doped ground-state from the 3-spin polaron, consisting of a singlet between a Cu spin and the spin of the doped hole occupying a coherent $x^2-y^2$ linear combination of the O orbitals surrounding it.\cite{Esk1988} However, it is important to note that a  3-spin polaron Bloch state and its ZRS counterpart are not orthogonal, but instead have a momentum dependent overlap which is large at antiferromagnetic Brillouin zone boundary. In particular, this is the case at  the lowest energy removal state located at ($\pi/2,\pi/2$), as confirmed by ARPES \cite{Pot1997, Kim1998}.

Even though easier to study numerically, the phase diagrams of these one-band models is still being debated. In particular it has not yet been demonstrated conclusively that they host high-temperature superconductivity in the thermodynamic limit, although they do show other relevant behavior, such as strong antiferromagnetic (AFM) correlations \cite{Now2012,Him1999} and the appearance of stripes. \cite{Hua2018} 

The discussion, so far, has completely ignored electron-phonon (e-ph)
coupling, following one branch of thought that assumes that purely
electronic Hamiltonians should suffice to describe high-temperature
superconductivity. This resulted from early ideas that e-ph coupling cannot
possibly be strong enough to drive such high critical temperatures,
and therefore the pairing must have a different ({\em i.e.} electronic)  origin.
On the other hand, 
multiple experiments have provided evidence that e-ph coupling is
important in understanding cuprates -- although this may be due to the interpretation of the experimental results in terms of models based on one-electron physics, where correlations play a minor roles. For example, Lanzara \textit{et al.}
investigated the quasiparticle dispersion of LSCO and Bi2212 at
different doping levels by ultrafast electron spectroscopy
\cite{Lan2001} and found a kink at around 50-80 meV, suggestive of
e-ph coupling to phonons with this energy. Shiina et al.
\cite{Shi1995} and Shimada et al. \cite{Shi1998} measured the
tunneling conductance of Bi$_2$Sr$_2$CaCu$_2$O$_8$ (BiSCCO) and
observed peaks in $d^2I/dV^2$ matching those in the phonon density of
states. Similar results were found using scanning tunneling microscopy
on BiSCCO \cite{Cas2008}. More recently, angle-resolved photoemission
spectroscopy on (closely related) 1D cuprate chains has revealed a
strong near-neighbor attraction \cite{Che2021} which was
explained as arising from longer-range e-ph coupling.
\cite{Wan2021}

This raises the question of what happens when e-ph coupling is added to the electronic Hamiltonians already mentioned above. These Hamiltonians are strongly correlated so their response to even a weak el-ph coupling could be quite different from what is expected in non-correlated systems. There have already been attempts to investigate the outcome using  the Hubbard-Holstein model, which adds the simplest possible e-ph coupling to the one-band Hubbard model. The results suggest a positive cooperation that may favors higher T$_c$ values in the presence of the Holstein e-ph coupling \cite{Gun2008,Mis2004,Mac2004}.

While this is an encouraging result, one important question is whether the Holstein model is a good description of the e-ph coupling in the complex perovskite structures. Our recent work shows that that is not always the case. \cite{Yam2020} Even when it is, an even deeper issue is how to properly gauge the strength of the e-ph coupling. This strength can be estimated for the bare hole, but that is {\em not} the fermion described by the one-band models. Instead, the latter fermion describes a ZRS or 3-spin polaron-like quasiparticle already strongly dressed by very local correlations, and which therefore has a rather low overlap with the bare  hole. This overlap can be thought of as a `coefficient of fractional parentage' or as a `quasiparticle weight' resulting from the projection of the three-band model onto the low-energy correlated manifold that is the basis of the one-band model. For example, within the ZRS picture, the effective $p$-$d$ AFM exchange that stabilizes the ZRS is of order 3 eV, as found by studying a small cluster using the three-band like model plus full on-site Coulomb atomic multiplet interactions.\cite{Esk1989} The calculated large energy splitting between the low-energy singlet and the high-energy triplet states for two holes in a CuO$_4$ cluster support a strongly reduced coefficient of fractional parentage. This was confirmed in studies of larger clusters, where the quasiparticle weight was found to be reduced to about 0.2 at $(\pi/2,\pi/2)$  \cite{Lau2011,Lau2011_2} (with a strong momentum dependence and even smaller values in other parts of the Brillouin zone).  We expect that this strong reduction must be affecting the strength of the quasiparticle-phonon coupling in a non-trivial way.

To investigate this issue, in this work we consider the hole-phonon coupling that arises in the simplified Emery model with spins at the Cu sites and doped holes moving on the O sublattice. The Hamiltonian that we use to describe it was proposed by Lau et al. \cite{Lau2011} and Ebrahimnejad et al. \cite{Ebr2014} as the strongly-correlated limit of the three-band model.  The quasiparticle of this model (in the absence of hole-phonon coupling) was studied both with exact diagonalization and with a variational method, and was shown to have a dispersion in good agreement with that measured experimentally. Furthermore, the result is robust in the sense that the dispersion has the correct shape without need for fine-tuning of parameters, unlike  in one-band models.

Our starting point for studying the hole-phonon coupling is to add to the original three-band model an e-ph coupling of Peierls type\cite{Bar1970,Su1979,Hee1988}, which modulates the magnitudes of the $t_{pd}$ and $t_{pp}$ hopping of the bare holes between Cu and O and  between O orbitals, respectively,  when the lighter O ions oscillate.
  The strength of this `bare' coupling can be estimated by hand, as discussed below. As already mentioned, to avoid the complications of dealing with this really complex model,  here we study its strongly-correlated limit when a single hole is doped in the system. Thus, the electronic part is identical to that of  model derived and studied by Lau et al. \cite{Lau2011}. We derive the additional e-ph coupling by tracing the effects of the bare Peierls coupling in this  strongly-correlated, low energy manifold. As detailed below, this turns out to be quite complicated and certainly not Holstein-like. We then  generalize  the variational method used in Ebrahimnejad et al. \cite{Ebr2014} to   study the effect of this projected e-ph coupling on the dispersion of the quasiparticle, and find that it has a  very small effect on the quasiparticle's effective mass. This confirms that the  `projected' coupling is weak, as a direct consequence of the considerable  renormalization of the coupling of the correlation-dressed quasiparticle when compared to that of a bare hole.

The work is organized as follows: In Section II we introduce the three-band model and its Peierls coupling to phonons, and then derive its strongly-correlated limit and discuss the resulting hole-phonon couplings. In Section III we briefly discuss the underlying ideas of the variational method we use, with technical details relegated to appendixes. Section IV contains our results, and Section V has an extended discussion of our findings.

\section{\label{method}Model}

As illustrated in Fig. \ref{lat_for_H}, the  three-band Emery model of the CuO$_2$ plane includes Cu 3d$_{x^2-y^2}$ orbitals and the O 2p$_\sigma$ ligand orbitals, arranged on a Lieb lattice. The corresponding Hamiltonian reads:
\begin{align*}
H_{3B}=&T_{pp}+T_{pd} +\Delta\sum_{\ell,\epsilon,\sigma} n_{\ell+\epsilon,\sigma}
\\
&+U_{pp}\sum_{\ell,\epsilon} 
n_{\ell+\epsilon,\uparrow}n_{\ell+\epsilon,\downarrow}+U_{dd}\sum_{\ell} 
n_{\ell,\uparrow}n_{\ell,\downarrow}
\numberthis\label{3-band}
\end{align*}
Here, $\ell$ are the site labels for $d$ orbitals, and $\ell+\epsilon$ are the labels for $p$ orbitals, where $\epsilon \in \{\epsilon_x,\epsilon_y \}$ point to  $x$  and $y$ ligand O (see Fig. \ref{lat_for_H}). We denote by $d_{\ell,\sigma}, p_{\ell+\epsilon, \sigma}$ the corresponding {\it hole} annihilation operators; the hole number operators are $n_{\ell,\sigma}=d^\dagger_{\ell,\sigma}d_{\ell,\sigma}$,  $n_{\ell+\epsilon,\sigma} = p^\dagger_{\ell+\epsilon, \sigma}p_{\ell+\epsilon, \sigma}$.

With this notation and the orbitals chosen as shown in Fig. \ref{lat_for_H}, the hopping terms are:
\begin{align*}
	T_{pd}=t_{pd}\sum_{\ell, \epsilon,\sigma}
	(-p^{\dagger}_{\ell+\epsilon,\sigma}+
	p^{\dagger}_{\ell-\epsilon,\sigma})
	d_{\ell,\sigma}+\text{h.c.}
\numberthis\label{tpd}
\end{align*}
and
\begin{align*}
T_{pp}=t_{pp}&\sum_{\ell,\sigma}[p^{\dagger}_{\ell+\epsilon_x,\sigma}
(-p_{\ell+\epsilon_y,\sigma}+p_{\ell-y+\epsilon_y,\sigma}\\
&-p_{\ell+x-y+\epsilon_y,\sigma}
+p_{\ell+x+\epsilon_y,\sigma})
+\text{h.c.}]\\
-t'_{pp}&\sum_{\ell,\epsilon,\sigma}
(p^{\dagger}_{\ell-\epsilon,\sigma}
+p^{\dagger}_{\ell+3\epsilon,\sigma})
p_{\ell+\epsilon,\sigma}
\numberthis\label{tpp}
\end{align*}
where $t_{pd}>0, t_{pp}>0$ are the magnitudes of nearest-neighbour (NN) hopping between $pd$ and $pp$ orbitals, respectively,  and $t'_{pp}>0$ is the magnitude of next nearest-neighbour (NNN) hopping between two O bridged by a Cu.
The other three terms of $H_{3B}$ describe the charge-transfer energy and the on-site Hubbard repulsion at O and Cu sites, respectively. 

The simplest (Einstein) description of the optical phonon modes  is obtained assuming that each O ion oscillates along its ligand bond about its equilibrium position, in between the two much heavier (effectively immobile) Cu neighbours. This leads to the Einstein optical phonon Hamiltonian:
\begin{align*}
H_{ph}=\Omega\sum_{\ell,\epsilon} b^{\dagger}_{\ell+\epsilon}b_{\ell+\epsilon}
\numberthis\label{Hph}
\end{align*}
where $b^{\dagger}_{\ell+\epsilon}$ creates a phonon at the O located at $\ell+\epsilon$ (we set $\hbar=1$ throughout).

To obtain the hole-lattice coupling, we note that the $pd$ and $pp$ hoppings are modulated by the motion of the O involved in the process.  $U_{pp}$ and $U_{dd}$ are not modulated by the O motion, hence they do not contribute  to the hole-lattice coupling. The charge transfer $\Delta$ is affected  because the modulation of $t_{pd}$ changes the covalence of the $p$-$d$ bonds,\cite{Lan2001_2,Lor1995}  however we expect this to be a smaller effect and we ignore it in the following. To lowest order, the hopping between equilibrium positions described in  Eqs. (\ref{tpd}), (\ref{tpp}) is therefore  supplemented by small corrections proportional to the displacements out of equilibrium. They read:
\begin{align*}
	H^{pd}_{h-ph}=g\sum_{\ell, \epsilon,\sigma} 
	\big[&p^{\dagger}_{\ell+\epsilon,\sigma}d_{\ell,\sigma}
\hat{u}_{\ell+\epsilon}
	+p^{\dagger}_{\ell-\epsilon,\sigma}d_{\ell,\sigma}
	\hat{u}_{\ell-\epsilon}
	+\text{h.c.}\big]
\end{align*}
\begin{align*}
	H^{pp}_{h-ph}=g_{pp}\sum_{
	\ell,\sigma}
	\left[p^\dagger_{\ell+\epsilon_x,\sigma}
	p_{\ell+\epsilon_y,\sigma}(\hat{u}_{\ell+\epsilon_x}+\hat{u}_{\ell+\epsilon_y})+h.c +\dots\right]
\end{align*}
where  $\hat{u}_{\ell+\epsilon} \equiv b^\dagger_{\ell+\epsilon}+b_{\ell+\epsilon}$, and for the $pp$ part we wrote explicitly only the terms for one bond; $\dots$ stand in for similar terms for the other three bonds. We ignore the modulation to the longer range hoping  $t'_{pp}$  because it is of much smaller magnitude than these two.

The hole-phonon coupling is, then:
\begin{align*}
	H_{h-ph}=H^{pd}_{h-ph}+H^{pp}_{h-ph},
\end{align*}
and the total Hamiltonian is $H = H_{3B}+H_{ph}+T_{h-ph}$. 

Generally accepted values for the parameters of the three-band model
are: $t_{pd}=1.3$eV, $t_{pp}=0.65$eV, $t_{pp}'\approx0.38$eV,
$\Delta=3.6$eV, $U_{pp}=4$eV, $U_{dd}=10.6$eV. \cite{Oga2008,
  Lau2011}.
For the phonons, we use a typical optical energy of $\Omega=0.090$ eV.
\cite{pin1994} To estimate the strength of the electron-phonon
couplings, we assume that the hopping integrals obey Harrison's
rules\cite{harbook}, {\it eg.} $t_{pd}\propto d^{-7/2}$ where $d=a/2+\delta u$ is
the $pd$ distance, $\delta u$ being its displacement from equilibrium.
Taylor expanding to first order in $\delta u
=\sqrt{\frac{\hbar}{2m\Omega}}\hat{u}$ leads to
$g=\frac{7}{a}\sqrt{\frac{\hbar}{2m\Omega}}t_{pd}=0.091$eV. Similarly, starting
from $t_{pp}\propto d^{-3}$ we find $g_{pp}=\frac{3}{a}\sqrt{\frac{\hbar}{2m\Omega}}t_{pp}=0.020$eV. We note that while these couplings are close to those derived in Ref.  \onlinecite{hor2005}, they are about 3-4 times smaller than the Holstein  coupling used to explain the lifetime of the quasiparticle peak as measured in ARPES, when  starting from a one-band $tJ$ model  \cite{Mis2004}. We discuss below the difference in the coupling strengths for a Peierls vs Holstein model, which may well account for this difference. Nevertheless, for illustration purposes, we will also generate results for our model with both $g$, $ g_{pp}$ couplings increased by a factor of three (within the limit of our computational power),  to see the effect of such an increase on the results.

This model is too complicated to study for our purposes. To make
further progress, we note that spectroscopic studies \cite{Fuj1987}
show that the doped holes reside primarily on the oxygen sites, to
avoid the costly Hubbard $U_{dd}$ which is the largest energy in the
problem. To mimic this behaviour more simply, we let $U_{dd}\rightarrow \infty$ so as
to prevent doubly occupied Cu $d$-sites. This leads to an easier
problem with spins at singly occupied Cu sites,  and
doped holes moving on the O sublattice and interacting with the Cu
spins.

To obtain the model describing this simplified problem,  we use perturbation theory to find the effective Hamiltonian projected on the lowest manifold with singly occupied Cu states. More specifically, for the undoped ground state we have to project on the manifold spanned by the states $\prod_\ell d^\dagger_{\ell,\sigma_{\ell}}\ket{0}$ with each $\sigma_\ell =\uparrow, \downarrow$. To 4th order perturbation theory in $T_{pd}$, this leads to the antiferromagnetic superexchange $J_{dd} = \frac{8t_{pd}^4}{\Delta^2(U_{pp}+2\Delta)}= 0.157$ eV between neighbor Cu spins. The electron-phonon coupling will slightly renormalize the value of $J_{dd}$ but symmetry prevents the appearance of terms linear in the $\hat{u}_{\ell+\epsilon}$ operators, and for consistency we ignore non-linear couplings.

Before continuing, it is important to note that for finite $U_{dd}$ there is a second contribution to $J_{dd}$ (and all the other effective parameters derived below). In Appendix \ref{ap-vals} we present the corresponding formulae for a finite $U_{dd}$ and explain why we choose to ignore these terms, {\em i.e.} to use the $U_{dd} \rightarrow \infty$ values. 

We make the further approximation of treating this exchange as Ising instead of Heisenberg. The accuracy of this approximation was validated in previous work \cite{Ebr2016} and will be briefly reviewed below. Hence, the effective Hamiltonian describing interactions between Cu spins is taken to be:
\begin{align*}
	H_{dd}=J_{dd} \sum_{\langle\ell,\ell'\rangle}{S}^z_\ell{S}^z_{\ell'} (1- n_{(\ell+\ell')/2})
	\label{H_dd}
	\numberthis
\end{align*}
The last factor enforces the fact that this superexchange vanishes on the bond hosting the doped hole. 
From now on, we set $J_{dd}\approx 0.150$eV as our unit of energy.

To obtain the full effective Hamiltonian when there is an additional hole doped into an O orbital, we use 2nd order perturbation theory to project onto the manifold spanned by the states $p^\dagger_{l_0+\epsilon_0,\sigma_0}\prod_\ell d^\dagger_{\ell,\sigma_{\ell}}\ket{0}$. The calculation is similar to that in Ref. \onlinecite{Lau2011} but with the coupling to phonons now included. The effective Hamiltonian describing the doped hole is found to be:
\begin{align*}
	H_{\text{eff}}&=H_{dd}+ \Delta
	\sum_{\ell,\epsilon} n_{\ell+\epsilon,\sigma}
	 +T_{pp}+H_{pd}+T_{sw} \\ &+H_{ph}
	+H^{pp}_{h-ph}+H^{pd}_{h-ph}+H^{sw}_{h-ph}
	\label{Heff}
	\numberthis
\end{align*}
where the first line is the effective model derived in Ref. \onlinecite{Lau2011}. The new terms describing  the hole-phonon coupling are on the second line.

Specifically, besides the superexchange $H_{dd}$ between Cu spins on the bonds not hosting the hole, and the  hopping $T_{pp}$ of the doped hole in the undisturbed O lattice (supplemented by the Peierls modulation described above in $H^{pp}_{h-ph}$), there are two additional terms describing the interactions between the hole and its neighbor Cu sites. The first is an AFM exchange between their spins. In the absence of hole-phonon coupling, this is:
\begin{align*}
  H_{pd}=J_{pd} \sum_{l, \epsilon}\vec{S}_l\cdot\vec{S}_{l\pm \epsilon}
	\label{Hpd}
	\numberthis
\end{align*}
where $J_{pd}=\frac{2t^2_{pd}}{\Delta+U_{pp}}=2.84J_{dd} $. Note that this is a Heisenberg coupling, which allows the doped hole at $\ell \pm \epsilon$ to flip its spin if the Cu spin at $\ell$ also flips. For clarity, we note that this is {\em not} the previously mentioned $p$-$d$ exchange of order 3eV that stabilizes the ZRS. That much larger exchange arises between the Cu spin and the spin of a hole on the $x^2-y^2$ `molecular' orbital of surrounding O, and can be shown to be a linear combination of this `atomic' $J_{pd}$ as well as the $t_{sw}$ term defined below. The O motion modulates $t_{pd}$ and therefore the magnitude of this $J_{pd}$ exchange. To linear order, this gives an additional hole-phonon coupling:
\begin{align*}
  H^{pd}_{h-ph}=&g_{pd}\sum_{\ell,\epsilon}\big[\mp
	(S^+_\ell S^-_{\ell\pm\epsilon}+S^-_\ell S^+_{\ell\pm \epsilon})
	(b^\dagger_{\ell\pm\epsilon}+b_{\ell\pm \epsilon})\\
	&\pm(\frac{1}{2}-2S^z_\ell S^z_{\ell\pm\epsilon})
	(b^\dagger_{\ell\pm \epsilon}+b_{\ell\pm\epsilon})
	\big]
	\label{Hpd-hph}
	\numberthis
\end{align*}
where $g_{pd}=\frac{gt_{pd}}{\Delta+U_{pp}}+\frac{gt_{pd}}{\Delta+\Omega+U_{pp}}= 0.197J_{dd}$. 

Thus, $H_{pd}+H^{pd}_{h-ph}$ describe the processes where the hole from Cu$_\ell$ does a virtual hop to its neighbor O$_{\ell \pm \epsilon}$ that hosts the doped hole, and then one of the two holes returns to Cu$_{\ell}$. The sister process where the doped hole hops from O$_{\ell \pm \epsilon}$ to Cu$_\ell$ and then one hole returns to the same O$_{\ell \pm \epsilon}$ is forbidden if $U_{dd}\rightarrow \infty$, but for a finite $U_{dd}$ it simply renormalizes the values of $J_{pd}, g_{pd}$ as discussed in Appendix \ref{ap-vals}.

The second additional term is the `swap' term which describes the related processes where if the doped hole is initially at $\ell+\epsilon$, then the hole of either Cu$_\ell$ or of Cu$_{\ell + 2\epsilon}$ hops to any of its three other O neighbors (labelled $\ell+\epsilon+\eta$) followed by the doped hole from O$_l$ moving to that Cu. Effectively, the doped hole has hoped from $O_{\ell+\epsilon}$ to $O_{\ell+\epsilon+\eta}$ while swapping its spin with that of the Cu$_{l_{\epsilon,\eta}}$ that neighbors both of these O. If the O are at their equilibrium positions, this gives:
\begin{align*}
  T_{sw}=-t_{sw} \sum_{l, \epsilon, \eta \atop \sigma, \sigma'} \xi_{\eta}p^\dagger_{\ell+\epsilon+\eta,\sigma} p_{\ell+\epsilon,\sigma'}|\sigma'\rangle_{l_{\epsilon,\eta}l_{\epsilon,\eta}}\langle\sigma|
	\label{Hsw}
	\numberthis
\end{align*}
where $t_{sw} = \frac{t_{pd}^2}{\Delta}= 2.98J_{dd}$. Here, $\eta$ is either one of the four $\delta$ vectors (see Fig. 1) connecting to NN O, in which case $\xi_{\delta}=\pm 1$ if $\delta$ is oriented at 45$^\circ$ above/below the horizontal; or $\eta = \pm 2\epsilon$ points to the two NNN O bridged through a Cu, in which case $\xi_{\pm 2\epsilon}=1$. $|\sigma\rangle_{l_{\epsilon,\eta}}$ indicates the spin of Cu$_{l_{\epsilon,\eta}}$ which is NN to both $O_{\ell+\epsilon}$ and $O_{\ell+\epsilon+\eta}$, {\it i.e.} $\ell_{\epsilon,\eta}\equiv\ell+\epsilon+\frac{\eta\cdot\epsilon}{|\eta\cdot\epsilon|}\epsilon$.

Finally, displacements of either of these two O  will modulate the strength of this swap term, leading to another  hole-phonon linear coupling term:
\begin{widetext}
\begin{align*}
  H^{sw}_{h-ph}=&\sum_{\ell,\epsilon,\eta \atop \sigma,\sigma'}
	\big[\big(g_i \xi^d_\eta b^\dagger_{\ell+\epsilon}+
	g_f\frac{\epsilon\cdot\eta}{|\epsilon\cdot\eta|}
	b^\dagger_{\ell+\epsilon+\eta}\big)
	p^\dagger_{\ell+\epsilon+\eta,\sigma} p_{\ell+\epsilon,\sigma'}+\big(g_i \xi^d_\eta b_{\ell+\epsilon}+
	g_f\frac{\epsilon\cdot\eta}{|\epsilon\cdot\eta|}
	b_{\ell+\epsilon+\eta}\big)
	p^\dagger_{\ell+\epsilon,\sigma} p_{\ell+\epsilon+\eta,\sigma'}\big]|\sigma'\rangle_{l_{\epsilon,\eta}l_{\epsilon,\eta}}\langle\sigma|
	\label{Hsw-hph}
	\numberthis
\end{align*}
\end{widetext}
where $g_i\equiv\frac{gt_{pd}}{\Delta}= 0.208J_{dd}$ and $g_f\equiv\frac{gt_{pd}}{\Delta+\Omega}= 0.203J_{dd}$ and $ \xi^d_\eta=\pm1$ corresponding to the positive/negative overlap between the lobes of $p_{\ell+\epsilon+\eta}$ and $d_{\ell_{\epsilon,\eta}}$.

The effective one-hole Hamiltonian $H_{\rm eff}$ of Eq. (\ref{Heff}) describes a complex polaronic problem where the hole can (i) emit/absorb  magnons from the magnetic background of the Cu spins (spin off-diagonal terms in $H_{pd}$ and $T_{sw}$); (ii)  emit/absorb phonons ($H^{pp}_{h-ph}$ and the spin-diagonal terms in $H^{pd}_{h-ph}+H^{sw}_{h-ph}$); and (iii) simultaneously emit/absorb both magnons and phonons (through the spin off-diagonal parts of $H^{pd}_{h-ph}+H^{sw}_{h-ph}$).

Alternatively, the coupling to the bosons can be characterized as having both a diagonal (so-called $g({\bf q})$) component where the hole does not change its position during the absorption/emission of bosons (coupling terms arising from $H_{pd}+H^{pd}_{h-ph}$) and an off-diagonal (so called $g({\bf k,q})$) component where the hole moves while the bosons are emitted/absorbed (coupling terms arising from $H^{pp}_{h-ph}+T_{sw}+H^{sw}_{h-ph}$).

\section{\label{method}Variational Approximation}

Because our goal is to understand how the coupling to phonons affects the quasiparticle's mass, we need to first find the quasiparticle dispersion $E_{\bf k}$. This can be extracted as the lowest-energy pole in the spectrum of the single-hole propagators:
\begin{align*}
G_{\alpha\beta}({\bf k},\omega)
=\mel{{\bf k},\alpha,\uparrow}{{\hat G}(\omega)}{{\bf k},\beta,\uparrow}
\end{align*}
Here ${\hat G}(\omega)\equiv[\omega-H_{\rm eff}+i\eta]^{-1}$ is the retarded rezolvent of $H_{\rm eff}$ and
$$|{\bf k}, \beta, \uparrow\rangle\equiv\frac{1}{\sqrt{N}}\sum_{\ell+\epsilon \in O_\beta}e^{i{\bf k}\cdot {\bf R}_{\ell+\epsilon}}p^\dagger_{\ell+\epsilon,\uparrow}\ket{\mbox{AFM}}$$
is the Bloch state describing the hole located on equivalent sites $\beta$ of the O sublattice while the spin background is in its undoped ground-state $|\mbox{AFM}\rangle$. Without loss of generality, we assume that the doped hole is injected with spin-$\uparrow$. $N\rightarrow \infty$ is the number of unit cells.

\begin{figure}[t]
	\centering
	\includegraphics[width=0.5\columnwidth]{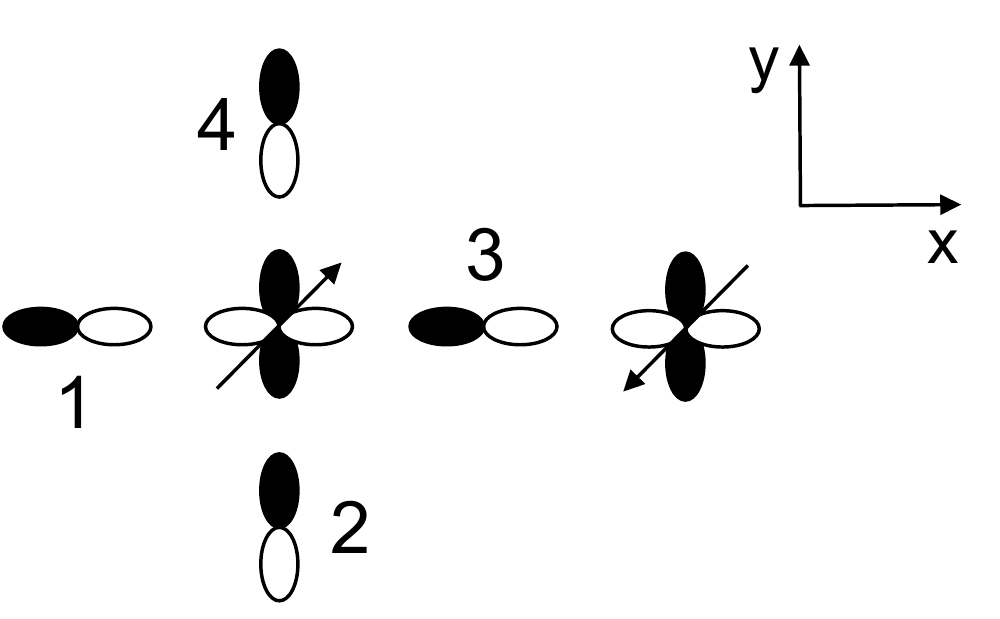}
	\caption{\label{label_p_orb} Our choice of the magnetic unit cell, and the labeling used for the four distinct O sites.}
\end{figure}

As already mentioned, we take $|\mbox{AFM}\rangle\rightarrow|\mbox{N\'eel}\rangle$ to be the N\'eel state stabilized by the Ising superexchange $J_{dd}$ (the justification is briefly discussed below). As a result, there are two inequivalent Cu sites (one with spin-up, one with spin-down) and four inequivalent O sites in the unit cell, see Fig. \ref{label_p_orb}, thus $\alpha,\beta = 1,2,3,4$.

We use the variational Momentum Average Approximation (MA), to solve for these  $G_{\alpha\beta}({\bf k},\omega)$ Green's functions. This approach generalizes the work of Refs. \onlinecite{Ebr2014,Ebr2016} where the same problem -- without coupling to phonons -- was solved by a similar variational approximation. There, it was revealed that the magnon cloud accompanying the hole is rather small, with up to about three magnons, although the quantitative differences between constraining the variational space to allow only 2 vs. 3 magnons, were small. It is important to emphasize that these are magnons (spin-flips) emitted and absorbed by the hole in its immediate vicinity, as it moves through and interacts with the magnetic background. The ground-state of the full Heisenberg model would also host background spin fluctuations, whose existence is independent of the presence of the doped hole. In Refs. \onlinecite{Ebr2014,Ebr2016} it was shown that  {\em insofar as the dynamics of the quasiparticle is concerned}, these background spin-fluctuations can be ignored in this three-band  model because they have little effect on it; this is because the time scale over which the quasiparticle propagates is significantly faster than that over which background spin-fluctuations occur. Turning 'off' these irrelevant (for our purposes) background spin-fluctuations is achieved by replacing the Heisenberg exchange with the Ising one, as done in Eq. (\ref{H_dd}).

We build on these results by restricting the variational space to allow for up to two magnons, but also a phonon cloud. We also continue to ignore the background spin-fluctuations. This latter approximation can only be justified {\it a posteriori}, if it turns out that the mass of the new quasiparticle is not significantly heavier than that found in the absence of hole-phonon coupling. If, instead, the new quasiparticle is much heavier (slower), then the time scale over which it propagates could become comparable to that over which background spin-fluctuations act, and they would need to be included in the calculation. As we show below, it turns out that the results fall in the former category.

We now review the main idea and the other approximations involved, with technical details relegated to Appendix \ref{Idea_MA}. 
To calculate the one-hole propagators, we divide the effective Hamiltonian as 
$	H_{\text{eff}}=H_0+H_\text{h-b}$, 
where we group all terms which change either the number of magnons (by flipping a Cu spin) and/or the number of phonons into  $H_\text{h-b}$, describing the coupling of the hole to both species of bosons. All other terms (that conserve the numbers of bosons) are part of $H_0$. We then use Dyson's identity $\hat{G}(\omega) = \hat{G}_0(\omega) + \hat{G}(\omega)  H_{\rm h-b} \hat{G}_0(\omega)$, where $\hat{G}_0(\omega)$ is the rezolvent for $H_0$, to link the $G_{\alpha\beta}({\bf k},\omega)$ propagators to generalized propagators that have either a magnon, or a phonon, or one of each bosons. The equations of motion for these new propagators, obtained by applying Dyson's identity again, link them to other generalized propagators with more phonons and/or magnons; and so on,  generating an infinite hierarchy of coupled equations.

We use variational principles to select which ones of these generalized propagators are large at low energies because their bosonic configuration has a high overlap with the ground-state. These propagators are kept in the hierarchy of coupled equations, while all other (smaller) generalized propagators are set to zero. This simplifies the system of coupled equations so that they can be solved numerically. The quality of the choice made for the variational space can be verified by adding more bosonic configurations, to see if they change the results. 

As already mentioned, previous work found that good accuracy is achieved by limiting the magnon configurations to have up to two magnons (in the absence of hole-phonon coupling) \cite{Ebr2014,Ebr2016}. Interestingly, a similar problem was also solved to see the effects of the hole-phonon coupling in the non-interacting limit ($U_{dd}=U_{pp}=0$), when there are no magnons. For lattices with similar structure in both 1D, 2D and 3D, it was found that the phonon cloud remains small spatially (even though it can host many phonons if the coupling is strong, {\it i.e.} the local deformation can be significant), see Refs. \onlinecite{Mol2016,Mol2017,Yam2020}.

We combine these results, and therefore constrain the boson configurations to allow up to two magnons plus a cloud of phonons on a single O site. We keep all configurations consistent with the distance between the most distant bosons being below a cutoff, whose value is increased until convergence is achieved. This defines the system of coupled equations that needs to be solved to find $G_{\alpha\beta}({\bf k},\omega)$, from which we extract the quasiparticle dispersion. More technical details are in  Appendix \ref{Idea_MA}.

\section{\label{sec:results}Results}

\begin{figure}[t]
	\centering
	\includegraphics[width=1\columnwidth]{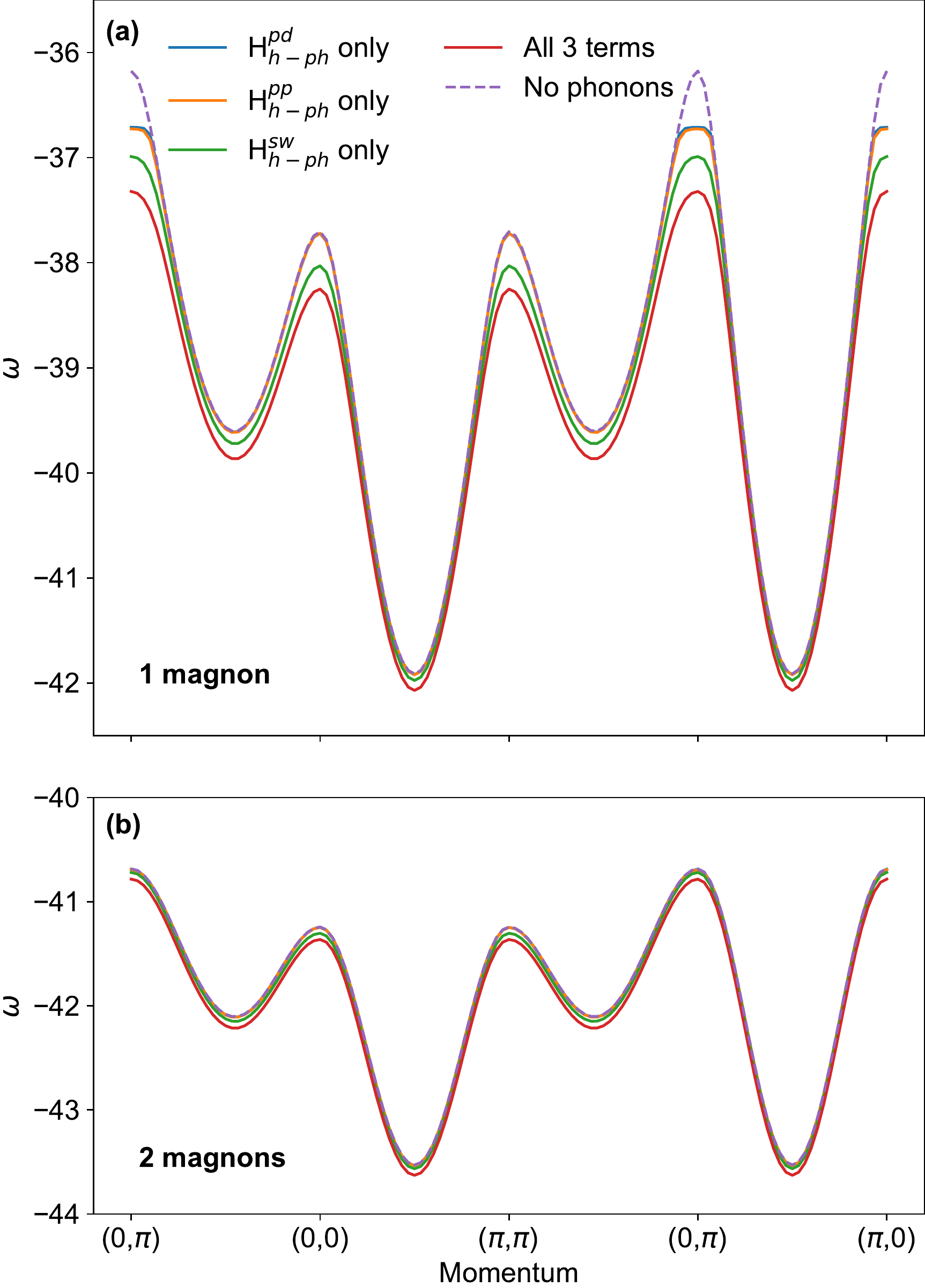}
	\caption{\label{1m2m_diff_term} Quasiparticle dispersion $E({\bf k})$ along high symmetry lines in the magnetic Brillouin zone. The variational calculation is limited to up to (a) 1 magnon and (b) 2 magnons, plus a phonon cloud. The dashed line labeled 'no phonons' shows the results when the hole-phonon coupling vanishes. The three lines labeled $H^{pd}_{h-ph}, H^{pp}_{h-ph}$ and $H^{sw}_{h-ph}$ show the results when only that particular hole-phonon coupling term is included. Finally, the line labeled 'all 3 terms' corresponds to including all three hole-phonon coupling terms.}
%
\end{figure}

We begin by plotting in Figure \ref{1m2m_diff_term} the quasiparticle dispersion $E({\bf k})$ along various cuts in the Brillouin zone. The results labelled `No phonons' correspond to setting the hole-phonon coupling to zero, and agree with the results obtained in previous work \cite{Ebr2014,Ebr2016} in the one magnon (top panel) and two magnons (bottom panel) variational spaces. Clearly, a bigger magnon cloud leads to a slower quasiparticle with a narrower bandwidth, but the same overall shape of the dispersion. In particular, there is a nearly isotropic minimum around the ground-state located at ${\pi\over 2a}(1,1)$, in agreement with experimental data \cite{Dam2003}.

Adding coupling to phonons has a very small effect on the dispersion. The additional curves show results when only one of the three possible hole-phonon couplings  (originating from $T_{pd}, T_{sw}$ and $H_{pd}$, respectively) are turned on, as well as their total combined effect.

\begin{table}[t]
  \centering
  \begin{tabular}{ |c|c|c| } 
 \hline
$m^*/m_e$   &  $(0,0)-(\pi,\pi)$ & $(0,\pi)-(\pi,0)$\\ 
 \hline
 1 magnon & 1.156 & 0.921\\ 
 \hline
 1 magnon + phonons & 1.193 & 0.941 \\ 
 \hline
2 magnons & 1.620 & 1.277 \\ 
 \hline
 2 magnons + phonons & 1.640 & 1.281 \\
 \hline
\end{tabular}
  \caption{\label{tb1}Effective mass (in units of the bare electron mass $m_e$) calculated along $(0,0)-(\pi,\pi)$ and $(0,\pi)-(\pi,0)$ directions. The ground state momentum is $k_{gs}=(\frac{\pi}{2},\frac{\pi}{2})$. These results correspond to our calculated values of the hole-phonon coupling, with all three hole-phonon terms included. We used $a=1.9\AA$ for the lattice constant. }
 \end{table}

Next,  we calculate the quasiparticle effective mass along the two symmetry lines $(0,0)-(\pi,\pi)$  and $(0,\pi)-(\pi,0)$ from the dispersion near the ground-state minimum. The results are summarized in Table \ref{tb1}. First, we note that the effective masses are comparable to the free electron mass, in line with  experimental measurements\cite{Pad2005,Ram2015,Yel2008}. Second, it is important to emphasize that the effective mass $m^*$ in the absence of coupling to phonons is already significantly heavier than the bare band mass $m_b$ of a hole on the O sublattice, in the absence of coupling to the Cu spins (no correlations). One rough estimate of the mass enhancement due solely to correlations is the ratio of the bare bandwidth $8t_{pp}$ and the quasiparticle bandwidth of $2-3J_{dd}$, see Fig. \ref{1m2m_diff_term}(b). For our parameters, this gives $m^*/m_b\sim 13$. Another estimate is $m^*/m_b= 1/Z\sim 5$, where the quasiparticle weight (before coupling to phonons)  was found to be $Z\sim 0.2$ near the ground-state momentum $(\pi/2,\pi/2)$.\cite{Lau2011,Ebr2014} The second estimate is  likely  more accurate, given that the first one assumes bands whose dispersion arises only from nearest-neighbor hopping. In any event, it is clear that there is significant mass enhancement due to correlations, before turning on the electron-phonon coupling.

  Table \ref{tb1} shows that there is only a very minor additional increase of the effective mass when the coupling to phonons is added. The change is on the order of very few percent, suggesting that by this measure, this electron-phonon coupling is weak. This is not surprising, considering that the various hole-phonon couplings $g$ are order of magnitude smaller than the hole-magnon couplings controlled by $t_{sw}$ and $J_{pd}$  (we revisit this issue below).

Furthermore, the relative change in the effective mass  decreases for  the larger (two-magnon) variational space. At first sight this is not expected, because the `bare' (in terms of phonons) 2-magnon quasiparticle bandwidth is narrower, which would suggest a larger effective coupling to the phonons in this case. However, this is an oversimplified argument, based on Holstein-like couplings, and is known to not necessarily hold for the more complicated Peierls-like couplings \cite{Mar2010}. It also ignores the fact that the more dressed quasiparticle has a smaller `coefficient of fractional parentage' and thus a smaller effective coupling to phonons. We also note that terms in the hole-boson Hamiltonian that create both a phonon and a magnon, cannot act on one-magnon configurations unless two-magnon configurations are allowed. The effect of these additional terms may account for the decrease, although we could not fully disentangle these contributions, despite our best efforts. What we can say with confidence is that most of the effective mass increase comes from the hole-phonon modulation of the $T_{sw}$ term, not from those of $T_{pp}$ or $J_{pd}$.

\begin{table}[b]
  \centering
  \begin{tabular}{ |c|c|c| } 
 \hline
$m^*/m_e$   &  $(0,0)-(\pi,\pi)$ & $(0,\pi)-(\pi,0)$\\ 
 \hline
 1 magnon & 1.156 & 0.921 \\ 
 \hline
 1 magnon + phonons & 1.526 & 1.102\\ 
 \hline
2 magnons & 1.620 & 1.277 \\ 
 \hline
 2 magnons + phonons & 1.832 & 1.328 \\
 \hline
\end{tabular}
  \caption{\label{tb2}Same as Table \ref{tb1} except all hole-phonon couplings are increased by a factor of three, while keeping all other parameters unchanged.}
 \end{table}

We conclude that for this value of electron-phonon coupling, coupling to the phonons has  a  very minor influence on the quasiparticle mass.

To verify that phonons can have a less trivial effect, we repeat the calculation for an electron-phonon coupling that is three times larger than the values derived above, in other words we replace $H_{\rm h-b} \rightarrow 3 H_{\rm h-b}$ while keeping everything else unchanged. The corresponding results are shown in Table \ref{tb2}. The percentage increase is more significant in this case especially along the $(0,0)-(\pi,\pi)$ cut, confirming that phonons can affect the effective mass, as expected. However,  the overall trends are the same, in particular we again find a much smaller percentage increase in the larger variational space suggesting that fully converged results would find an increase of at most $\approx 10\%$ even for this stronger coupling.

\begin{figure}[t]
	\centering
	\includegraphics[width=0.65\columnwidth]{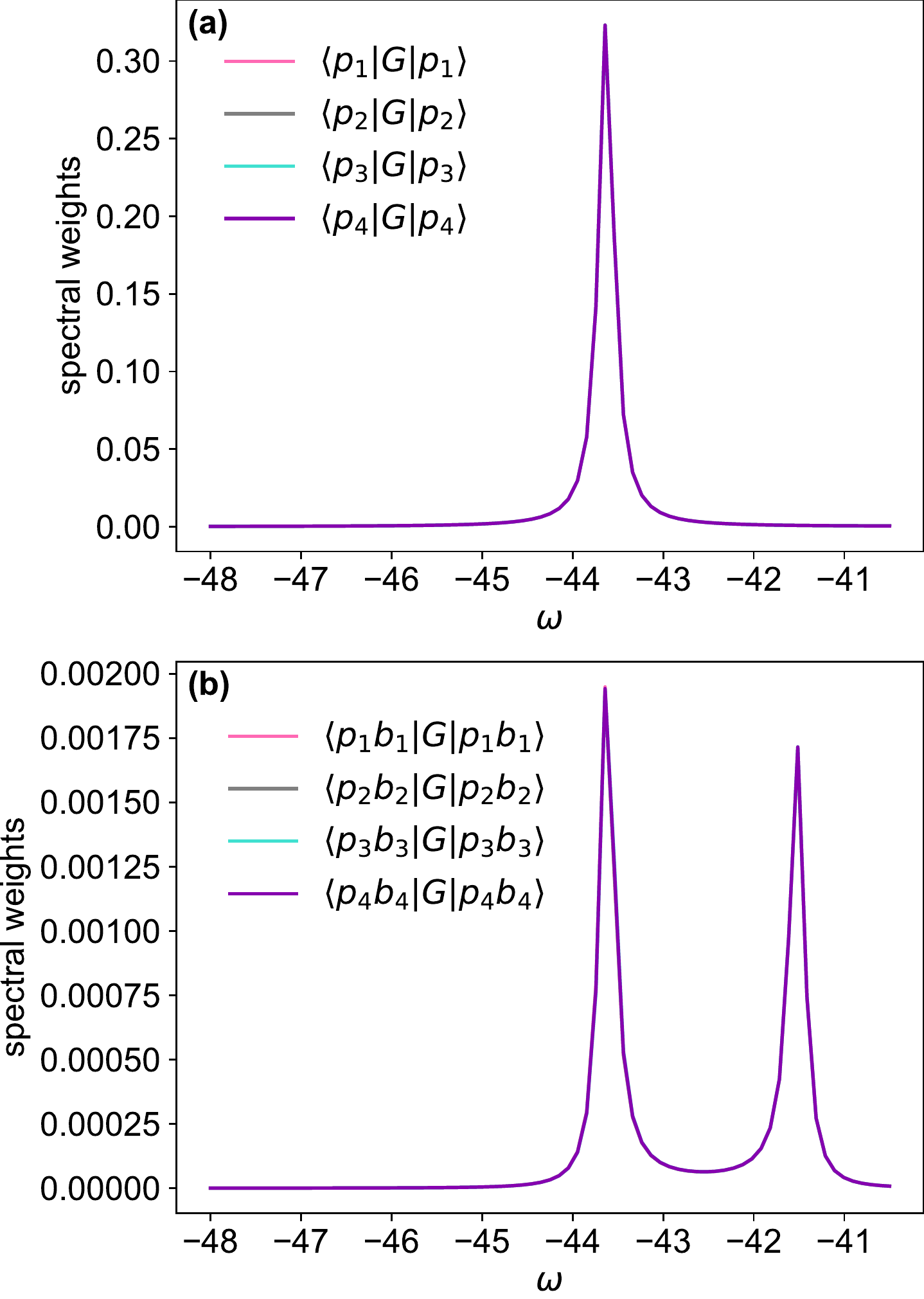}
	\caption{\label{p_pb_2m} At physical coupling and on ground state momentum $k=(\frac{\pi}{2},\frac{\pi}{2})$, (a) spectral weights of $\mel{p_i}{G}{p_i}$ and (b) approximations to the spectral weights of $\mel{p_ib_i}{G}{p_ib_i}$, where $i=1,\dots,4$. Same parameters are used as FIG.\ref{1m2m_diff_term}.}
\end{figure}

To further analyze the coupling to phonons, we revert to the original values of the electron-phonon couplings and plot in Fig. \ref{p_pb_2m}(a)  the spectral weights $A_\alpha({\bf k}_{\rm gs}, \omega)= - {1\over \pi}G_{\alpha\alpha}({\bf k}_{\rm gs},\omega)$ for $\alpha=1,..,4$. As expected, the quasiparticle weight is the same for all 4 O sublattices. In Fig. \ref{p_pb_2m}(b), we plot the ${\bf k}={\bf k}_{\rm gs}$ spectral weights associated with the generalized propagators $\langle {\bf k},\alpha,1,\uparrow|{\hat G}(\omega)|{\bf k},\alpha,1,\uparrow\rangle$, with $\alpha=1,..,4$, where now
  $$|{\bf k}, \alpha, 1, \uparrow\rangle\equiv\frac{1}{\sqrt{N}}\sum_{\ell+\epsilon \in O_\alpha}e^{i{\bf k}\cdot {\bf R}_{\ell+\epsilon}}p^\dagger_{\ell+\epsilon,\uparrow}b^\dagger_{\ell+\epsilon}\ket{\mbox{AFM}}$$
is the Bloch state that also has 1 phonon at the same O site where the doped hole resides.

We see that the latter spectral weight is about 150 times smaller, {\em i.e.}  the overlap $|\langle GS| {\bf k}_{\rm gs},\alpha,\uparrow\rangle|^2$ between the quasiparticle ground-state and the hole-only Bloch state is roughly 150 times larger than the overlap $|\langle GS| {\bf k}_{\rm gs},\alpha,1, \uparrow\rangle|^2$ of the ground-state with the Bloch state where a phonon accompanies the hole. Again, this is a confirmation that this electron-phonon coupling is weak (the probability of exciting phonons in the quasiparticle cloud is rather low).

\begin{figure}[t]
	\centering
	\includegraphics[width=1\columnwidth]{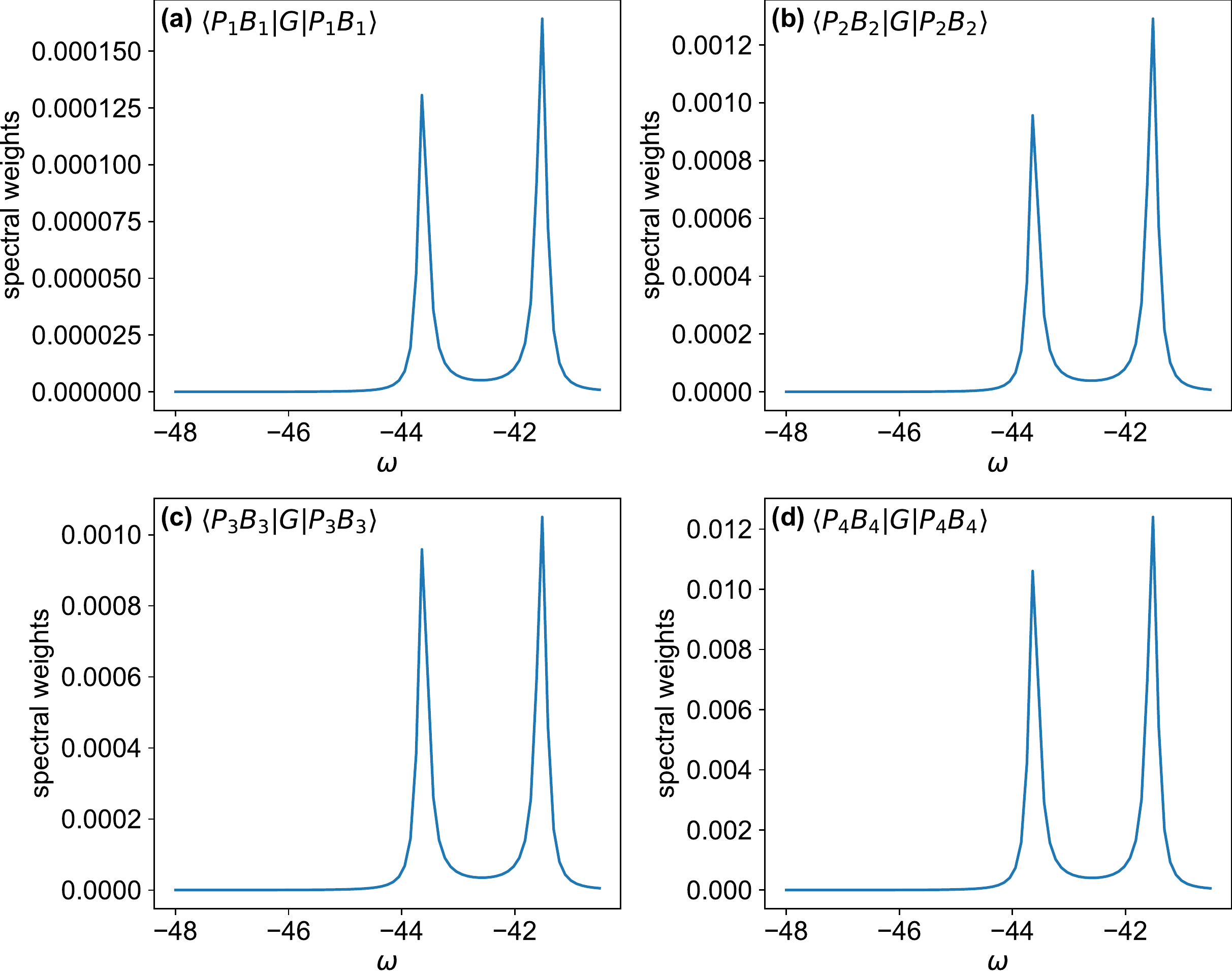}
	\caption{\label{PB_2m} Spectral weights  $-{1\over \pi} Im \mel{P_\alpha B_\alpha}{{\hat G}(\omega)}{P_\alpha B_\alpha}$ (see text for details) for physical coupling and at the ground state momentum ${\bf k}_{\rm gs}=(\frac{\pi}{2},\frac{\pi}{2})$,  for $\alpha=1$ to $4$ ((a) to (d)). All parameters are the same as in Fig. \ref{1m2m_diff_term}.}
\end{figure}

However, we can also project the one-phonon states in a different basis, consistent with the expected local symmetry of the quasiparticle. We define the new hole operators:
\begin{align*}
P_1&=\frac{1}{\sqrt{4}}(p_1-p_2-p_3+p_4)\\
P_2&=\frac{1}{\sqrt{2}}(-p_1-p_3)\\
P_3&=\frac{1}{\sqrt{2}}(p_2+p_4)\\
P_4&=\frac{1}{\sqrt{4}}(p_1+p_2-p_3-p_4)
\label{sym_state}
\numberthis
\end{align*}
where we use the short-hand notation $p_\alpha$ for the operators associated with the hole being on the 4 O sites of the unit cell, see Fig. 2. As a result, $P_1$ has local $s$-symmetry, $P_2$ and $P_3$ have the $p_x$ and $p_y$ symmetries respectively, and $P_4$ has local $x^2-y^2$ symmetry. We also define similar `molecular' deformations with various local symmetries, and associated the phonon operators  $B_i$ to describe them. We then define one-phonon Bloch states $|P_\alpha B_\alpha\rangle \equiv \frac{1}{\sqrt{N}}\sum_{\ell+\epsilon \in O_\alpha}e^{i{\bf k}_{gs}\cdot {\bf R}_{\ell+\epsilon}}P^\dagger_{\ell+\epsilon,\uparrow}B^\dagger_{\ell+\epsilon}\ket{\mbox{AFM}}$ and calculate their corresponding spectral weights. The results are shown in Fig. \ref{PB_2m}, where the four panels correspond to the four values of $\alpha$.

We now see that for three of these symmetries $\alpha=1,...,3$, the overlap with the quasiparticle ground-state wavefunction has gone down by yet another order of magnitude. However, for $\alpha=4$, the overlap is now roughly 6 times larger than before.  If we use the three-times larger coupling, the corresponding spectral weight is 10 times larger than in Fig. \ref{PB_2m}(d) (not shown).

On the one hand, this confirms again that the quasiparticle has a Zhang-Rice singlet type of nature \cite{Zha1988,Esk1988}, with a local $x^2-y^2$ symmetry. 
More importantly, this shows that when characterizing the strength of the hole-phonon coupling in such complex lattices, it makes a significant difference if one quantifies the strength of the coupling to individual sites versus the coupling to `molecular'-like orbitals consistent with the expected local symmetry: for the same problem, the latter can be order(s) of magnitude stronger than the former. We believe that this accounts for some of the discrepancy in literature between the values of Peierls couplings to individual sites (as derived here) vs. Holstein-like coupling to `molecular' orbitals mimicking states in a single band. This issue is discussed in more detail next.

\section{\label{sec:conclusion}Discussion}

In this work, our starting point is the fact that upon doping of a parent cuprate layer, the main charge propagation channel is through the O $2p$ holes,  because strong correlations (large $U_{dd}$) suppress Cu $3d^8$ configurations. Projecting out these $3d^8$ states, we obtain an effective Hamiltonian whose new parameters $J_{dd}$, $J_{pd}$ and $t_{sw}$ depend on the value of $t_{pd}$. 
We then added the change in the O$2p$-Cu$3d$ hoping integral $t_{pd}$ with the interatomic distance, which is the largest contribution to the hole-lattice coupling (complemented by the associated modulation of $t_{pp}$, which is included in our model for completeness). Projecting this onto the lower-energy manifold of states with Cu $3d^9$ configurations allowed us to find the corresponding hole-phonon coupling in the strongly correlated limit of the three-band model. The resulting terms are conceptually simple but mathematically fairly involved, and describe the modulation of the  $J_{pd}$ and $t_{sw}$ processes due to the motion of the involved O ions, in addition to the afore-mentioned modulation of $T_{pp}$.

We then used a well established variational method to study the influence of this hole-phonon coupling on the dispersion of the quasiparticle, and found it to  be very small: the dispersion is little changed by coupling to phonons, when compared to that of the already strongly magnon-dressed quasiparticle (spin-polaron) obtained in the absence of phonon coupling\cite{Ebr2014,Ebr2016}.

This is a positive result, insofar as the alternative ({\it i.e.} a quasiparticle made much heavier by the hole-lattice coupling) would definitely be detrimental to the possibility of finding high-temperature superconductivity at finite dopings. We  emphasize that on their own, the results presented here do not mean that the hole-lattice coupling is irrelevant to this problem. This coupling affects not only the quasiparticle dispersion but also the  hole-hole effective interaction mediated through boson exchange. If addition of phonon exchange to the magnon exchange   will turn out to boost the hole-hole effective attraction \cite{Mol2019} (as is the case in the simpler Holstein-Hubbard model) then hole-lattice coupling could play a significant role in driving high-temperature superconductivity in the cuprates, even if it may not be its primary driver. The calculation of this effective attraction in the presence of both magnons and phonons is a complicated matter, which will be postponed for future work.

Returning to our main result, namely that within our model and with the approximations we used, the hole-phonon coupling  has little consequences on the quasiparticle's dispersion: On one hand, this is not a huge surprise given that the  $g\approx 0.6J_{dd}$  we estimate from the modulation of $t_{pd}$, is significantly smaller  than the $J_{pd}$ and $t_{sw}$ energy scales, suggesting that dressing by magnons is dominant over dressing by phonons. Moreover, the actual couplings to phonons in our effective model come from the modulations of $J_{pd}$ and $t_{sw}$ due to their $t_{pd}$ dependence, and  are even smaller, $g_{pd}$, $g_i$, $g_f \approx 0.2J_{dd}$. It is unclear how to properly define an effective coupling $\lambda$ for this complex model, but if we use the Holstein formula then, {\em e.g.},  $g_{pd}^2/\Omega W \sim 0.06$, where $W \sim J_{dd}$ is half the bandwidth of the spin polaron dispersion. Adding three such contributions gives a total $\lambda < 0.2$ that is very small, consistent with the minor effect on the effective mass. 

On the other hand, if we consider the original $g\sim \Omega$, then $g^2/\Omega W \sim 0.6$ suggests a much stronger coupling, consistent with  previous work which claimed that in the underdoped limit,   $\lambda \sim 0.5- 1$. \cite{Lan2001, Lan2001_2, Ros2005, Gun2008}  

Part of the answer for this discrepancy has already been pointed out at the end of the previous section, namely that the coupling to the `molecular'-like orbital involved in the Zhang-Rice singlet (which could be a rough proxy for the Holstein coupling strength in a one-band model) is larger than the coupling to individual O sites. Indeed, an increase by a factor of 2-3 of this `molecular $g_{pd}$' -- consistent with the observed order of magnitude increase  for the probability to excite a `molecular' breathing-mode phonon $B_4$  -- suffices to increase the corresponding  $\lambda$ by a factor of 4-9, to become of order 1. 

However, such arguments ignore the very important fact that coupling of an already dressed quasiparticle (our spin polaron, or the 'fermion' of the one-band models) to phonons is substantially smaller than the coupling of a bare hole to the same phonons, because of the coefficient of fractional parentage defined by the overlap between the quasiparticle and the bare hole eigenstates. As shown in previous work,\cite{Lau2011} this overlap is of order $Z\sim 0.2$ near $(\pi/2,\pi/2)$,  suggesting a $1/Z^2\sim 25$ times smaller effective coupling of the quasiparticle to the phonons. As a result, even a very strong bare hole-phonon  coupling can be  reduced to a very weak quasiparticle-phonon  coupling. We note that this is in qualitative agreement with Ref. \onlinecite{Gun2008}, although in a rather different context.

These arguments illustrate the main lesson of our work, namely that the modeling of carrier-phonon coupling in highly correlated models, where the `carrier' is an already significantly-dressed quasiparticle because of correlations, is a subtle problem that needs to be considered very carefully. Our results suggest that it is dangerous to assume that the form of that coupling is  simple, and demonstrate that it is wrong to assume that the strength of that coupling equals the coupling of the bare carrier to phonons. The alternative to trying to guess the correct coupling for quasiparticle is to proceed like we did here, by starting from a more complex model describing the bare carriers subject to both correlations and coupling to the phonons. This requires more involved calculations but also much less uncertainty about reasonable parameters to be used. We propose to use the same approach to investigate next whether phonon-exchange can supplement the magnon-exchange to provide an enhanced glue for superconductivity in cuprates.

Furthermore, we aim  to provide a detailed description of how to treat electron-phonon coupling in systems which already have strongly dressed quasiparticles because of other interactions, thus generalizing this work beyond cuprates.

\acknowledgements

We acknowledge support from the Natural Sciences and Engineering Research Council of Canada (NSERC), the Stewart
Blusson Quantum Matter Institute (SBQMI) and the
Max-Planck-UBC-UTokyo Center for Quantum Materials.

\appendix
\section{The effective parameters}
\label{ap-vals}

If we keep the contribution from doubly-occupied Cu $3d$ states in the various perturbative expressions of the effective parameters, we find the expressions:
\begin{align*}
J_{dd}'=&2\left(\frac{4t^4_{pd}}{\Delta^2(U_{pp}+2\Delta)}+\frac{2t^4_{pd}}{\Delta^2U_{dd}}
\right)=0.240\text{eV}\\
t_{sw}'=&\frac{t_{pd}^2}{\Delta}+\frac{t_{pd}^2}{U_{dd}-\Delta}=0.710\text{eV}\\
J_{pd}'=&2\left(\frac{t_{pd}^2}{U_{pp}+\Delta}+\frac{t_{pd}^2}{U_{dd}-\Delta}\right)
=0.926\text{eV}\\
\end{align*}
If we take this  $J_{dd}'$ as our energy unit, we find $t'_{sw} = 2.96 J'_{dd}$  very close to the ratio $t_{sw}=2.95J_{dd}$ used in the main text, while  $J'_{pd} = 3.85 J'_{dd}$ is somewhat bigger than the corresponding ratio $J_{pd}=2.84J_{dd}$. The ratios are not hugely different because all processes acquire a second channel, with a contribution roughly equal to that of the first channel. The overall results will therefore remain roughly the same, apart from the increase in the energy scale from $J_{dd}=0.150$eV to $J_{dd}'=0.240$eV.

At first sight, this latter value appears to be problematic because it does not agree with the measured superexchange in cuprates, unlike the former. However, it is wrong to assume that the primed expressions are more `accurate' simply because they include the finite $U_{dd}$ contributions. In fact, it is clear that the validity of all these perturbative estimates is very questionable, considering that the ratio $t_{pd}/\Delta \approx 0.36$ is far from `very small', and so are the other relevant  ratios, too.

Thus, to obtain accurate estimates for these effective parameters one has to go to higher order in perturbation theory, which is an unpleasant prospect. Instead, we take the pragmatic approach to assume that these higher order corrections will renormalize the various effective parameters roughly similarly (for the same reasons as above), so that their ratios remain essentially unchanged. The  value for $J_{dd}$ energy unit is then chosen to be in agreement with experiment. Then end result is a Hamiltonian whose parameters are very close to the values $J_{dd}, t_{sw}, J_{pd}$ we used in our main text. 

%

\section{Technical details for the  Momentum Average (MA) Approximation \label{Idea_MA}}

As discussed in the main text, we  define the Green's functions $G_{\alpha\beta}(\bf{k},\omega) = \mel{k,\alpha,\uparrow}{\hat{G}(z)}{k,\beta,\uparrow}$ where
$
\ket{\bf{k},\beta,\uparrow}\equiv\frac{1}{\sqrt{N}}\sum_{\ell+ \epsilon \in O_\beta}e^{i\bf{k}\cdot\bf{R}_{\ell + \epsilon}}p^\dagger_{\ell + \epsilon,\uparrow}\ket{\rm{AFM}}
$
is a Bloch state associated with the hole occupying the $\beta=1,\dots,4$ O2$p$ orbitals  (see Fig. 2) where $R_{\ell+ \epsilon}$ is the location of the $p$ orbital of type $\beta$, the Cu spins are in their N\'eel order $|\rm{AFM}\rangle$ , $N \rightarrow \infty$ is the number of unit cells, $\hat{G}(z)=[ z - H_{\text{eff}}]^{-1}$ is the rezolvent of our effective Hamiltonian defined in Eq. (\ref{Heff}), and $z=\omega + i\eta$, where $\eta$ is a small artificial broadening.

To generate equations of motion, we use Dyson's identity $\hat{G}(z)=\hat{G}_0(z)+\hat{G}(z)H_{\rm{h-b}}\hat{G}_0(z)$ where we divide $H_{\text{eff}}=H_0+H_{\rm{h-b}}$, with $H_{\rm{h-b}}$ collecting all terms that change the number of bosons (magnons and/or phonons) while $H_0$ collects the terms that conserve the number of bosons. The resulting infinite hierarchy of equations of motion is simplified by only keeping the generalized propagators consistent with boson configurations included in our variational space. Specifically, if we limit the variational configurations to having a single magnon and/or a single one-site phonon cloud (the smallest choice), this is equivalent with only considering the hierarchy involving additional generalized propagators such as:
\begin{align*}
V_r\equiv \sum_{\ell}\frac{e^{i{\bf k\cdot R_{\ell}}}}{\sqrt{N}}&\bra{{\bf k},\alpha,\uparrow}\hat{G}(z)
p^\dagger_{\ell+r,\downarrow} S^+_{\ell}\ket{\rm{AFM}}\\
f^{s,n}_r\equiv\sum_{\ell}\frac{e^{i{\bf k\cdot R_{\ell}}}}{\sqrt{N}}&\bra{{\bf k},\alpha,\uparrow}\hat{G}(z)p^\dagger_{\ell+r,\uparrow}(b^\dagger_{\ell + s})^n
\ket{\rm{AFM}}\\
\tilde{f}^{s,n}_r\equiv \sum_{\ell}\frac{e^{i{\bf k \cdot R_{\ell}}}}{\sqrt{N}}
&\bra{{\bf k},\alpha,\uparrow}\hat{G}(z) p^\dagger_{\ell+r,\downarrow}(b^\dagger_{\ell + s})^nS^+_{\ell}\ket{\rm{AFM}}
\end{align*}
For simplicity of notation, the dependence of these propagators on ${\bf k}, z, \alpha$ is not written explicitly.

When we allow up to two magnons in the variational space, we include two more kinds of generalized Green functions:
\begin{align*}
W_{r,\xi}=\sum_{R_\ell}&\frac{e^{i\bf{k\cdot R_{\ell}}}}{\sqrt{N}}
\bra{\bf{k},\alpha,\uparrow}\hat{G}(z)p^\dagger_{\ell+r,\uparrow}
	S^+_{\ell}S^-_{\ell+\xi}
	\ket{\rm{AFM}}\\
\bar{\tilde{f}}^{s,n}_{r,\xi}=\sum_{R_\ell}&\frac{e^{i\bf{k\cdot R_\ell}}}
{\sqrt{N}}\bra{\bf{k},\alpha,\uparrow}\hat{G}
p^\dagger_{\ell+r,\uparrow}(b^\dagger_{\ell +s})^nS^+_{\ell}S^-_{\ell+\xi}
\ket{AFM}
\end{align*}

Even if we only keep these propagators in the hierarchy of equations of motion, while setting all other generalized propagators to zero, the result is still an infinite system of coupled equations of motion (albeit it with a much simpler structure than the exact one). We further truncate it by limiting the spatial relative distances $r,s,\xi$ to be less than a cutoff. The low-energy quasiparticle is a coherent state where the magnon+phonon clouds are bound to the hole, and thus these relative distances are rather small. Indeed, we find that the low-energy part of the spectrum converges fast with this distance cutoff, as well as with the cutoff $n\le N_{ph}$ defining the  maximum number of phonons allowed in the cloud. The results shown here are converged with respect to both of these cutoffs.

\bibliography{article}

\end{document}